\documentstyle[prb,aps,epsf]{revtex} \def\narrowtext{} \tighten \twocolumn
\input epsf.sty

\begin{document}
\draft

\title{The Relation of Neutron Incommensurability to Electronic Structure in
High Temperature Superconductors}
\author{M. R. Norman}
\address{
Materials Science Division, Argonne National Laboratory, 
Argonne, IL  60439}

\address{%
\begin{minipage}[t]{6.0in}
\begin{abstract}
The relation between the incommensurability observed in neutron scattering
experiments in bilayer cuprate superconductors and the electronic structure
is investigated.  It is found that the observed incommesurability pattern, as
well as its dependence on energy, can be well reproduced by electronic
dispersions motivated by angle resolved photoemission data.  The commensurate
resonance and its contribution to the superconducting condensation energy
are discussed in the context of these calculations.
\typeout{polish abstract}
\end{abstract}
\end{minipage}}

\maketitle
\narrowtext

\section{Introduction}

One of the more controversial topics in the field of high temperature cuprate
superconductivity is the origin of the incommensurability observed by neutron
scattering experiments.  The original explanation of this phenomenon was that it
was due to Fermi surface nesting\cite{FS}.  Later, it was proposed that it was
due to the formation of stripes\cite{TRANQ}.  These two explanations are so
different, one would expect that ways of differentiating them using existing
data should be possible.  In bilayer cuprates, the situation is even more
interesting, in that a commensurate resonance is seen, along with
incommensurability at energies off resonance\cite{MOOK2,ARAI}.

In this paper, the Fermi surface nesting approach is analyzed based on tight
binding energy dispersions motivated by angle resolved photoemission
spectroscopy (ARPES) studies, with the dynamic
susceptibility calculated using the random phase approximation (RPA).
Such tight binding dispersions are able to reproduce the
commensurate resonance, along with the incommensurability off resonance.  In
particular, the incommensurability wavevector is found to be energy dependent,
in good agreement with recent experimental results\cite{ARAI}.  This would
seem to argue against a stripes interpretation, where one would expect the
incommensurability to be energy independent.  Moreover, the
incommensurability in the RPA calculations is very sensitive to the
underlying electronic structure, which has implications for the position
and curvature of
the Fermi surface near the d-wave node.  As for the commensurate resonance,
it is a more robust feature of these calculations, though
its width in momentum space is sensitive to the location of the flat band
near $(\pi,0)$.  In support of earlier estimates \cite{SW,DZ,MOOK3},
the change in exchange energy between the normal
and superconducting states in these calculations is sufficient to account
for the superconducting condensation energy.

In Section II, the details of the computations are given.  In Section III,
the commensurate resonance is discussed in relation to the electronic structure,
with Section IV dealing with the same relation in regards to the
incommensurability off resonance.  Section V uses these calculations to
comment on the question of the lowering of the exchange energy in the
superconducting state.  Some conclusions are offered in Section VI.

\section{Calculational Details}

The methodology used here is essentially the same as that of
other groups\cite{LEVIN,SCALAPINO,LAVAGNA,BLEE}.  The only difference
is to use energy dispersions motivated by actual fits to photoemission
measurements.  As in these earlier studies, one first determines the
non-interacting susceptibility, which in the superconducting state
is composed of three terms\cite{SCHRIEFFER,SCALAPINO}:
\begin{eqnarray}
\lefteqn{\chi_0(q,\omega) =} \nonumber \\
& & \sum_k\{\frac{1}{2}(1+\frac{\epsilon_k\epsilon_{k+q}+\Delta_k\Delta_{k+q}}
{E_kE_{k+q}})\frac{f(E_{k+q})-f(E_k)}{\omega-(E_{k+q}-E_k)+i\delta} \nonumber \\
& & +\frac{1}{4}(1-\frac{\epsilon_k\epsilon_{k+q}+\Delta_k\Delta_{k+q}}
{E_kE_{k+q}})\frac{1-f(E_{k+q})-f(E_k)}{\omega+(E_{k+q}+E_k)
+i\delta} \nonumber \\
& & +\frac{1}{4}(1-\frac{\epsilon_k\epsilon_{k+q}+\Delta_k\Delta_{k+q}}
{E_kE_{k+q}})\frac{f(E_{k+q})+f(E_k)-1}{\omega-(E_{k+q}+E_k)+i\delta}\}
\end{eqnarray}
where $\epsilon_k$ is the dispersion, $\Delta_k$ the superconducting
gap, $E_k$ the quasiparticle energies ($\sqrt{\epsilon_k^2+\Delta_k^2}$),
and $f$ the Fermi function.  The three terms are due to quasiparticle
scattering, quasiparticle pair annihilation, and quasiparticle pair
creation, respectively.  At low temperatures, only the last process
contributes.  Note this equation makes a severe approximation of treating
the single particle spectral function as a delta function.
On the other hand, at low temperatures, photoemission data
indicate quasiparticle peaks at all ${\bf k}$ vectors near the Fermi
surface \cite{ADAM} (though along the d-wave node direction, there is still
some controversy\cite{VALLA}).  As the incoherent part of the spectral function
is unlikely to lead to sharp structure in ${\bf q}$ and $\omega$, these earlier
studies based on Eq.~1 are followed, but the reader should keep in mind
that there are certainly quantitative, and perhaps qualitative, differences
between this ``quasiparticle'' susceptibility
and the true one which warrant future investigation.

To evalulate Eq.~1, $i\delta$ is replaced by $i\Gamma$ in the energy
denominators.
The resulting well behaved function is then summed over a 400 by 400 mesh
in the Brillouin zone.  $\Gamma$, which is a crude representation of broadening
due to interactions and disorder, was taken to be 2.4 meV.  Smaller values of
$\Gamma$ simply lead to sharper structure in $\chi_0$ which requires a
denser ${\bf k}$ mesh in the sum.

The interacting susceptibility in the RPA is given by
\begin{equation}
\chi(q,\omega)=\frac{\chi_0(q,\omega)}{1-J(q)\chi_0(q,\omega)}
\end{equation}
were $J(q)$ is the spin-spin response function.  Many studies set $J(q)$ to
a constant, $J$, but in several
treatments\cite{LEVIN,BLEE}, $J(q)$
is assumed to be of the form $J_q=-J(\cos(q_xa)+\cos(q_ya))/2$ due to
superexchange between near-neighbor Cu sites, so calculations
were performed with both $J(q)=J$ and $J(q)=J_q$.  The
use of $J_q$ tends to suppress incommensurability since $J_q$ is
largest at the commensurate wavevector $(\pi,\pi)$.

For energy dispersions, a number of model dispersions present in the
literature\cite{RJ,NORM95,BLEE,SCHABEL} were analyzed.  These are based on tight
binding functions, and in the more sophisticated
models designed to fit photoemission data, they contain an expansion
up to a real space lattice vector of (2,2)\cite{RJ,NORM95,SCHABEL}.  The
first to be considered is one\cite{NORM95} which does a very good job of
modeling the normal state ARPES dispersion in $Bi_2Sr_2CaCu_2O_8$
(Bi2212)\cite{DING96}.
This is reproduced in
Table I.  Note that this model dispersion has no bilayer splitting, consistent
with its lack of experimental observation in Bi2212\cite{DING96}.  $J$ is then
chosen to yield a resonance at a particular energy.  A d-wave
energy gap is assumed of the form $\Delta_k=\Delta(\cos(k_x)-\cos(k_y))/2$.
An s-wave gap strongly reduces the tendency to obtain a commensurate resonance
because of the BCS coherence factors in Eq.~1\cite{DOUG,FONG}.

\section{Commensurate Resonance}

The first question addressed concerns the commensurate resonance, first
observed in $YBa_2Cu_3O_7$
(YBCO)\cite{ROSSAT,MOOK1,FONG,FONG2}, and recently in Bi2212\cite{MOOK4,KEIMER}.
The conditions under which a commensurate response is obtained for the
non-interacting susceptibility in a simple $t,t^\prime$ tight binding model is
given in Ref.~\onlinecite{LAVAGNA}, and are roughly obeyed for the more
sophisticated dispersions considered here.  For simplicity, consider first the
normal state.  Then, the condition reduces to whether the Fermi surface is
centered at $(0,0)$ or at $(\pi,\pi)$.  In the former case, the response
becomes incommensurate, basically because the Fermi surface and its shadow
image, displaced by a wavevector $(\pi,\pi)$, no longer intersect (the
intersection points are refered to as hot spots in the literature).  Once
a superconducting gap opens, there is a correction to this condition which
scales like $\Delta^2$ \cite{LAVAGNA}.  For the dispersion analyzed here, this
correction corresponds to about 5 meV.  That is, once the band at $(\pi,0)$
becomes more than 5 meV above the Fermi energy, the response becomes completely
incommensurate.  All models for YBCO, and most models for Bi2212,
are in favor of Fermi surfaces centered about $(\pi,\pi)$.  An exception is
a recent model by Chuang {\it et al} for Bi2212\cite{DESSAU}.  In this model,
the Fermi crossing is far enough away from $(\pi,0)$ along the $(0,0)-(\pi,0)$
line that an incommesurate response is expected, unlike what is observed
experimentally\cite{MOOK4}.

The non-interacting susceptibility having its maximum response at $(\pi,\pi)$
leads to a maximum at $(\pi,\pi)$ in the interacting susceptibility as well.
For a d-wave
gap, the coherence factors in Eq.~1 are maximal on the Fermi surface, due
the sign change of the gap under $(\pi,\pi)$ translation.  This leads to an
abrupt rise in Im$\chi_0$ above some threshold
energy value (twice the superconducting energy gap at the hot spots), which
is not present for the s-wave case.  By Kramers-Kronig
transformation, the abrupt rise causes a peak in Re$\chi_0$ \cite{AC}.
Therefore, if $J$ is large enough, a zero will occur in the real part of
the denominator of Eq.~2 for energies smaller than the threshold.  As
Im$\chi_0$ is small below the threshold, this leads to a
resonance like
behavior in Im$\chi$.  That is, in general, one expects a
resonance at a energy smaller than $2\Delta_{max}$.  These points are
illustrated in Fig.~1.
Note that all the peak structures in this plot
become more pronounced if the broadening, $\Gamma$, is reduced in size.

An interesting evolution of this resonance occurs as the chemical potential,
$\mu$, is varied.  Using the same model dispersion, $J$ was adjusted for each
$\mu$ in such a way that the maximum in Im$\chi$ occurs at 39 meV
(with $\Delta_{max}$ taken to be 35 meV).  In Fig.~2, a 2D contour plot is
shown at the resonance energy for various positions of the band at $(\pi,0)$,
$\epsilon(\pi,0)$.  As $\epsilon(\pi,0)$
approaches $\mu$ from below, the resonance broadens
in momentum space, and eventually transforms from a circular pattern to a
square-like pattern.  As the band crosses $\mu$, the resonance
pattern rotates to a diamond shape, and then far enough beyond, a completely
incommensurate response is achieved.  In agreement with the analytic
results on the $t,t^\prime$ model for Im$\chi_0$\cite{LAVAGNA}, the value
of $\epsilon(\pi,0)$ relative to $\mu$ where the crossover
to incommensurate behavior occurs is found to be proportional to $\Delta^2$.
The width of the resonance in momentum space is also sensitive to other
factors.  The above calculations assumed $J(q)=J$.
If $J(q)=J_q$ is used instead, the resonance would become narrower in $q$,
which is obvious from the $q$ dependence of $J_q$, $J_q$ being maximal at
$(\pi,\pi)$.  Also, the dispersion
used in the above calculation is based on normal state ARPES data.  In
the superconducting state, though, it is known that the dispersion becomes
flatter near $(\pi,0)$\cite{PRL97}, which acts to
broaden the resonance in momentum space.
From experiment, the resonance in Bi2212 is broader in
momentum space than in YBCO\cite{KEIMER}.  In the context here, this would
imply that for Bi2212 relative to YBCO: (1) the band at $(\pi,0)$ is closer to
the chemical potential, (2) the dispersion at $(\pi,0)$ is flatter, and/or
(3) $J(q)$ is flatter near $(\pi,\pi)$.
A quantitative comparison of experiment to theory will be discussed in
Section V.

At this stage, nothing has been said about the $q_z$ dependence of the
resonance.  Experimentally, the resonance only appears in the odd $(q_z=\pi)$
channel.  There are three ways this could occur in the context of the RPA
calculations.  First, $J(q)$ could be larger in the odd channel than the
even one due to interplane exchange\cite{LEVIN}.  In this case, the pronounced
gap in the even channel would be associated with the threshold energy
discussed earlier for Im$\chi_0$ (twice the energy gap at the hot spots).
Although experimentally, $J(q)$ is
larger in the odd channel\cite{HAYDEN}, the value of the interplane exchange
integral ($J_\perp$) is small enough that there would be qualitative
problems with fitting the even channel data.  For instance, for
both dispersions listed in Table I, given a value of $J$ needed to obtain
a resonance at 39 meV in the odd channel, a 20\% reduction in $J$
to simulate the even channel still results in a pronounced resonance.
An exception is the dispersion of Ref.~\onlinecite{SCHABEL}, where the
same analysis leads to no resonance in the even channel.  This occurs because
the peak in Re$\chi_0$ in Fig.~1 is very shallow.  For the same reason,
though, a value of $J$ much larger than experiment is needed ($\sim$ 1 eV) to
obtain a resonance in the odd channel.

The second way would be to
recognize that the odd channel corresponds to connecting bonding to
antibonding states in Eq.~1, whereas the even channel connects bonding to
bonding and antibonding to antibonding\cite{SCALAPINO}.  In this context,
bonding and antibonding refer to the bilayer splitting of the electronic
structure.  As mentioned above, there is little evidence for such
splitting from ARPES data in Bi2212, even in the superconducting
state\cite{DING96}
(although recently, this result has been challenged\cite{FENG}).  Still,
several calculations including bilayer splitting were analyzed.  In the first
case, dispersion two in Table I was taken to be the bonding band, and
the antibonding dispersion was gotten by adding a constant in such a way
that its Fermi surface crossing occured along the $(0,0)-(\pi,0)$ line.
In the second case, the bilayer split dispersion of Ref.~\onlinecite{SCHABEL}
was used.  In the first case, there was virtually no change in the results.
In the second case, there were some minor quantitative differences
(i.e., a pronounced resonance still existed in the even channel).
It should be remarked that although the evidence for bilayer
splitting is considerably greater in YBCO\cite{SCHABEL}, the
interpretation of ARPES data in this case is more controversial
due to the contribution from the chains, as well as surface related problems.

The third possiblity is that the $\langle\Delta_k\Delta_{k+Q}\rangle$
correlator in Eq.~1 only contributes
to the odd channel for some reason not apparent at the moment \cite{PWA}.
In this regard, it is interesting to note
that Janko \cite{JANKO} has recently predicted, based on thermodynamic
arguments, that the resonance will be strongly suppressed for fields along
the c-axis.  As Janko discusses, the likely source of this field dependence is
that the $\langle\Delta_k\Delta_{k+Q}\rangle$ correlator in Eq.~1 is sensitive
to phase coherence.
This has also been suggested to be the case in the context of interpreting
c-axis optical conductivity measurements \cite{IOFFE}.
This same phase coherence sensitivity might be linked to why the resonance only
appears in the odd channel.

\section{Incommensurate Response}

The next question concerns the incommensurate behavior off resonance.
Only in YBCO is
this known in detail\cite{MOOK2,ARAI}.  Incommensurate behavior is observed
in Bi2212 as well\cite{MOOK4}, but the 2D pattern of this in $q$ space,
as well as its energy dependence, is unknown at this
time.  Interestingly, in YBCO, the incommensurate wavevector depends on
energy\cite{ARAI}.  It basically has an ``hourglass" shape, with the
incommensurability wavevector approaching the commensurate value as the
resonance is approached from below, then again splitting out above the
resonance energy.  In the RPA calculations, although
incommensurability above the resonance energy is a robust result, being
related to the commensurate/incommensurate discussion of the previous
Section, the
incommensurability below the resonance energy is a different matter
altogether.  Incommensurability below resonance has been present in previous
calculations\cite{LEVIN2}, and the explanation for it within the RPA context
was given in a recent paper by Brinckmann and Lee\cite{BLEE}, with the
calculations presented here in agreement with their picture.  In the
superconducting state, the constant energy contours at low energies are very
elongated due to the large ratio of the Fermi velocity at the d-wave node to
the slope of the superconducting energy gap at the node.  This velocity ratio
has been experimentally determined by ARPES to be 20 for Bi2212\cite{MESOT},
and the same value has been extracted from low temperature thermal
conductivity measurements\cite{LOUIS}.  The latter measurements have also
determined the velocity ratio to be 14 in YBCO.
As shown by Brinckmann and Lee\cite{BLEE}, the incommensurability is due
to nesting between the energy contour about the d-wave node to the
same contour displaced by a wavevector $Q=(\pi,\pi)$, that is,
$E_k \sim \omega/2$, $E_{k+Q} \sim \omega/2$, where $\omega$ is the neutron
energy.  The wavevector is
incommensurate since the Fermi surface at the node is displaced away from the
$(\pi/2,\pi/2)$ points.

For the dispersion used in Fig.~2, though, no incommensurability is found
below the resonance energy, though incommensurability does occur above the
resonance energy.  This can be traced to the fact that for this
dispersion, the Fermi surface is too curved at the node.  This leads to
low energy contours which have a ``banana" shape, thus destroying nesting, as
illustrated in Fig.~3.
Moreover, for this dispersion, the Fermi crossing at the d-wave node corresponds
to a wavevector of $(0.37,0.37)\pi$, which would yield a larger
incommensurability than is typically observed.  Both of these
problems can be corrected if the Fermi wavevector is pushed closer to
the $(\pi/2,\pi/2)$ point.  This has the effect of reducing the Fermi
surface curvature, thus enhancing nesting, and also reducing the
magnitude of the incommensurability.  Reducing the Fermi velocity also aids
the nesting, but this is a smaller effect.

To investigate these points further,
several modifications to the dispersion used in Fig.~2\cite{NORM95} were made.
First, the very flat dispersion of the superconducting quasiparticle states near
$(\pi,0)$ was incorporated by setting $\epsilon(\pi,0)$ to -5 meV,
and then invoking the condition that the curvature of $\epsilon_k$ is 
zero along the $k_x$ and $k_y$ axes at this point.  To obtain an
incommensurability more relevant to experimental data, the Fermi wavevector
at the node was pushed out to $(0.41,0.41)\pi$.  At the same time,
the Fermi velocity was reduced from 1.6 eV$\AA$ to 1.0 eV$\AA$, but as
discussed above, this has a smaller influence on the results.  This
dispersion (two) is also listed in Table I,  with the resulting Fermi surface
and low energy contours shown in Fig.~3 as well.  Note the flatter energy
contours as compared to the previous dispersion.

In Fig.~4 (left panels), 2D contour plots are shown for this dispersion
at three different energies:  on
resonance, and 10 meV above and below resonance.  Note that the
incommensurability pattern below resonance has a striking ``baseball diamond"
shape, very similar to what has been observed experimentally in
YBCO\cite{MOOK2}.  This pattern was generated assuming $J(q)=J$.
If $J(q)=J_q$ is used instead (right panels), another maximum develops at the
commensurate wavevector, and is or is not the global maximum depending on
the particular energy.  In this context, a number of model dispersions
were analyzed.  Some exhibit completely commensurate behavior below
resonance, others completely incommensurate behavior, and others still
mixed behavior.  That is, in the RPA context, the incommensurability below
resonance is
very sensitive to the electronic strucuture, as well as to the momentum
dependence of $J(q)$.

In Fig.~5 (left panel), the wavevector along the $(\pi,\pi)$ direction
where Im$\chi$ is maximal is plotted versus energy,
using the same parameters as Fig.~4.  Note the distinct
``hourglass" shape of the pattern, which has recently been observed
experimentally in YBCO\cite{ARAI}.  The striking agreement of this pattern
with experiment is a strong argument in favor of an RPA-like interpretation
of the data.  On the other hand, the RPA calculations do suffer from
some quantitative problems.  In the right panel of Fig.~5, the intensity
at the wavevector where Im$\chi$ is maximal is plotted versus energy.
Note the extremely
sharp drop as one moves off resonance.  Experimentally, this drop is
less pronounced\cite{MOOK2}.

As suggested above, in the RPA context, the neutron scattering results are
a sensitive probe of the electronic structure.  This raises the question of
whether the incommensurability structure is in quantitative agreement with
ARPES results or not.  Unfortunately, all neutron scattering studies on this
issue but one have been done on YBCO.  ARPES results on YBCO are still somewhat
controversial because of surface related issues not present in Bi2212.
For instance, the tight binding dispersions proposed in
Ref.~\onlinecite{SCHABEL} do not support incommensurate behavior below the
resonance energy, again because of too strong a curvature of the Fermi
surface around the node (that is, the low energy contours are too curved
to support nesting).  Moreover, published ARPES results on YBCO
indicate a Fermi crossing along the node which would result in an
incommensurate wavevector which is too far displaced from $(\pi,\pi)$
relative to experiment.  It would, of course, be desirable in YBCO to
exploit the advent of the Scienta high momentum resolution detectors to
revisit this issue.

In Bi2212, the ARPES Fermi surface (in the vicinity of the d-wave node,
at least)
is better agreed upon.  As for neutron results, only one study has been
offered\cite{MOOK4}.  In that experiment, a rod of crystallites aligned along
the $(1,1)$ direction was measured.  This indicated an incommensurate
wavevector at low temperatures of around $(0.82,0.82)\pi$.  Note that the two
dimensional pattern of the structure is not known from these data.
Moreover, energy information is also not known, so there is always the
possibility that the incommensurability being observed is above resonance.
Lacking further data at this stage, let us assume that incommensurability is
indeed being observed below resonance.  Then, for the ARPES data to be
consistent with the neutron wavevector, the Fermi crossing along the
node would have to be at about $(0.40,0.40)\pi$.  This is close to a
recently reported value of $(0.39,0.39)\pi$ using a Scienta
detector \cite{VALLA}.  So,
the reported wavevectors from the ARPES and neutron experiments are certainly
within current error bars, and therefore consistent at this stage.
It would be highly desirable to: (1) have neutron data on the
incommensurability on single crystal
samples of Bi2212, and (2) to have more accurate measurements of both the
Fermi wavevector, and the curvature of the Fermi surface around the node,
from ARPES data, to further explore this point.

\section{Sum Rule and Condensation Energy}

Finally, the intriguing question of the relation of the commensurate resonance
to the superconducting condensation energy can be treated
very straightforwardly in the RPA context.  The advantage of these
calculations is that all wavevectors and energies are accounted for,
and therefore these calculations provide an important check to the ideas
proposed in Refs.~\onlinecite{SW} and \onlinecite{DZ}.

To begin with, $\chi$ has to be converted to units quoted in neutron
scatterering work.  This is achieved by multiplying Eq.~2 by the appropriate
matrix element.  For simplicity, consider the $zz$ matrix element, which is
$\sum_{\sigma}g^2\mu_B^2\langle\sigma|S_z|\sigma\rangle^2$.  For $g=2$, $S=1/2$,
this reduces to $2\mu_B^2$, as do the other two ($xx$, $yy$).
The sum rule \cite{SUM} can now be checked by summing Eq.~2 over the
zone, integrating $\omega$ out to the band edges, and multiplying by
$2\mu_B^2$.
The dispersion used in Fig.~5 is employed here, with a resonance at 39 meV
obtained by setting $J$ to 159 meV, a value comparable to experimental
estimates of $J$ in YBCO\cite{HAYDEN}.
The normal state is calculated by simply setting $\Delta$ to zero.  This is
a gross approximation, since it assumes quasiparticle states at T=0 for
the normal state, which is highly improbable.  In principle, a more
accurate represenation of the normal state could be simulated by increasing
$\Gamma$, but for simplicity, this is not done here.  For $J(q)=J_q$,
the sum is 1.62$\mu_B^2$ in the normal state, 1.64$\mu_B^2$ in the
superconducting state.  For $J(q)=J$, the sum is 1.93$\mu_B^2$ in the normal
state, 1.87$\mu_B^2$ in the superconducting state.
So, to within a few percent, the sum rule
is satisfied by the RPA calculations.  For a local moment system, we
would expect the value $\pi g^2 \mu_B^2 S(S+1)/3$\cite{HAYDEN,FONG2}, which for
$g=2$, $S=1/2$
reduces to $\pi$.  That is, the above values range from 52\% to 61\% of
the local moment result.  This reduction is to be expected, since Eq.~2 is
based on itinerant electrons.

A useful comparison to experiment is to integrate Im$\chi(\pi,\pi)$ over
energy.  Restricting to a 50 meV energy range, a value of 1.9$\mu_B^2$ per
plane is calculated, compared to an experimental value of 0.95$\mu_B^2$ in
Bi2212 and 0.8$\mu_B^2$ in YBCO\cite{KEIMER}.  This somewhat large
overshoot \cite{MORR}
is reduced when looking at the local susceptibility ($q$ integrated).
The maximum (at the resonance energy) per plane from the above
calculation is 9.8$\mu_B^2$/eV for $J(q)=J_q$, 14.9$\mu_B^2$/eV for $J(q)=J$.
This is comparable to the 9.5$\mu_B^2$/eV value quoted for underdoped
YBCO\cite{MOOK2}.  10 meV below resonance, the numbers are 1.1 and 2.2,
respectively, compared to an experimental value of 2.5\cite{MOOK2}.
As noted in the previous Section, for any given calculation, the
intensity appears to drop off below resonance faster than experiment.
Another useful comparison is to look at the full width half maximum of the
resonance in $q$ space.  The calculated values are
0.34$\AA^{-1}$ for $J(q)=J$, 0.23$\AA^{-1}$ for $J(q)=J_q$.  This is
to be compared to experimental values of 0.52$\AA^{-1}$ for Bi2212 and
0.25$\AA^{-1}$ for YBCO \cite{KEIMER}.

As noted in earlier work\cite{SW,DZ,MOOK3}, the exchange energy contribution
to the free energy in the t-J model, denoted as $E_X$, is obtained by
multipliying Im$\chi$ as defined in Eqs.~1 and 2 by $-3J_q/2\pi$, integrating
over
energy, and summing over the zone.  Its contribution to the condensation
energy is $E_X^{NS} - E_X^{SC}$, where $NS$ represents an extrapolation of
the normal state to zero temperature.  For the case considered above, a value
per plane of 28K is found if $J(q)=J_q$ is used, 59K if $J(q)=J$.
That is, within the t-J model context, the exchange energy is indeed
lowered in the superconducting state, with the calculated values somewhat
larger than those based on the data\cite{DZ,MOOK3}.  The advantage of the
current calculation is that this difference can be looked at as a function
of $q$ and $\omega$.  In Fig.~6a, the zone sum of $J_q$Im$\chi_q$
is plotted
as a function of $\omega$ ($J(q)=J$) for normal and superconducting states.
Note that the two merge at $2\Delta_{max}$, that is, the energy difference is
confined to energies below this.  In Fig.~6b, the $\omega$
integrated quantity is plotted versus $q$.  Note the contribution near
$(\pi,\pi)$ due to the resonance.  In addition, a normal state contribution
at low $q$ is removed in the superconducting state.  If $J(q)=J_q$ is
used instead (Fig.~6d), the low $q$ structure in the
normal state is suppressed, which is why the condensation energy is half
that of the $J(q)=J$ case.  This additional low $q$ structure is likely not
relevant to experiment since there is no evidence for it, and it is
doubtful whether the true $J(q)$ is large at small wavevectors.

The implications of Fig.~6 is clear.  The dominant contribution to the
exchange energy part of the condensation energy is due to the resonance.
This in support of previous work\cite{DZ,MOOK3}.
Moreover, as noted before\cite{DZ,MOOK3}, the estimated value, such as from
the above calculation, is more than sufficient to account
for experiment, as the total condensation energy has been estimated to be
only 3K per copper oxide plane from specific heat data for optimal doped
YBCO\cite{LORAM}.
(It should be mentioned that within the t-J context, if the exchange energy
is lowered in the superconducting state, then it is expected that the kinetic
energy would increase \cite{COND}.)
It is also interesting to remark that the RPA calculations appear to be
in greater agreement with the idea of Ref.~\onlinecite{DZ} (that the
resonance dominates the exchange energy difference) than related
calculations based on the spin-fermion model\cite{CHUB}, despite
qualitatively similar physics.  In the latter model, the merger noted in
connection with Fig.~6a occurs at much higher energies (of
order $J$).

Finally, some comments in regards to the nature of the resonance mode are
in order.
The quantum numbers of the resonance correspond to an excited
triplet ($S=1$) pair with center-of-mass momentum $Q=(\pi,\pi)$, since the
BCS ground state is $S=0$, $Q=0$, and the resonance is seen by spin flip
scattering.  On the other hand, the question of whether
the mode is actually a particle-particle mode is a more delicate
question, due to particle-hole mixing in the superconducting
state\cite{DEMLER}.  The RPA calculation assumes that the underlying action
is in the particle-hole channel.  Is there any experimental support for this?
As Demler and Zhang point out\cite{DEMLER}, this question can only be
indirectly answered by neutron scattering, as the neutrons only couple to
the particle-hole channel.  Their argument is that since the resonance only
appears below $T_c$, and since particle-hole mixing occurs below $T_c$, then
a particle-particle mode would only become visible in neutron scattering
below $T_c$,
in agreement with experiment.  Of course, in the RPA calculations presented
here, the mode also appears only in the superconducting state, since it is
a consequence of the BCS coherence factors in Eq.~1 (that is, the mode is
best not thought of as just a spin-wave mode).  Now, although it is true
that a spectral gap opens up above $T_c$ in underdoped materials (the
pseudogap), the RPA calculations are based on quasiparticle states, which
only appear below $T_c$\cite{PHENOM}.  So, in that sense, the Demler-Zhang
argument does not necessarily resolve this issue.

On the other hand, angle resolved photoemission does not have the same
restriction as neutron scattering.  Strong arguments have been made that the
the dramatic change in the spectral lineshape observed in such data below $T_c$
are a direct consequence of the interaction of the electrons with the
resonance mode\cite{PRL97,PRL99,AC}.  This comes from the dressing of the
electron
propagator by the resonance.  Note that unlike neutrons, there is nothing
restricting this coupling to be in the particle-hole channel.  That is, one
expects that if the mode were particle-particle in nature, the dominant
coupling to the self-energy of the electrons would be in the particle-particle
channel.  Let us now think about the simple limit that the mode energy
goes to zero.  Then, the dispersion of the higher binding energy feature
seen in ARPES data (the so-called hump) will have a dispersion given by
solving a very simple 2 by 2 secular equation\cite{KAMPF}.  In the
particle-hole case, the diagonal elements will be $\epsilon_k$ and
$\epsilon_{k+Q}$, where $Q=(\pi,\pi)$.  In the particle-particle case, the
second element would become $-\epsilon_{-k+Q}$ instead.  To first approximation
in both cases, the off-diagonal elements, denoted as $\Delta_U$, are taken to
be constants.  In Fig.~7, the dispersions obtained from both secular
equations are plotted using dispersion one of Table I
for $\epsilon_k$, for a typical value of $\Delta_U$ (100 meV).
For the particle-hole case, the resulting dispersion is very similar to
what is observed in photoemission\cite{MARSHALL,PRL99}.  But, the
particle-particle case has no resemblance at all to the data.

Now, this argument does not definitively rule out a particle-particle
explanation for the mode, since it is conceivable that the dominant coupling
of the mode to the electrons could still be in the particle-hole channel
because of the interaction vertices.  Still, the above argument is certainly 
very suggestive of a particle-hole origin for the mode, and would also more
naturally explain how the higher binding energy feature crosses
over to the Mott insulating gap as the doping is
reduced\cite{MARSHALL,LAUGHLIN,PRL99}.

\section{Conclusions}

In summary, this paper has shown that the RPA treatment of the dynamic
susceptibility gives very useful insight into neutron scattering data in
the cuprate superconductors.  In such a framework, the neutron data are
a sensitive probe of the underlying electronic structure.  Using dispersions
motivated by angle resolved photoemission data, a natural explanation is
found for the magnetic resonance observed by neutrons, as well as the
incommensurability seen off resonance.  Moreover, these calculations are
in support of previous suggestions\cite{DZ,MOOK3} that the resonance mode
provides the dominant contribution to the change in the exchange energy
between the normal and superconducting states.  The advantage of the RPA
calculations is that they provide quantitative information on all of these
issues.  In particular, the current study suggests that the two momentum
resolved probes used for the cuprates, ARPES and neutrons, are strongly
related to one another, and are consistent within current experimental
error bars.  With the advent of higher momentum resolution detectors in
ARPES, and large enough samples for neutron studies, this connection in
the future can be studied with much greater precision, especially in the case
of Bi2212.

\acknowledgments

The author would like to thank Herb Mook for suggesting this project.  He
would also like to thank Juan Carlos Campuzano, Helen Fretwell, and
Adam Kaminski for discussions concerning photoemission data.
This work was supported by the U.S. Dept. of Energy, Basic Energy 
Sciences, under Contract No. W-31-109-ENG-38.

\begin{table}
\caption{Tight binding dispersions based on fitting ARPES data.
The first two columns list the coefficient, $c_i$, of each term (eV), that is
$\epsilon(\vec k) = \sum c_i \eta_i(\vec k)$, with ``one" a previous
fit to normal state ARPES data\protect\cite{NORM95}, and ``two" a
modified fit as discussed in the text.  The last column lists the basis
functions (the lattice constant $a$ is set to unity).}
\begin{tabular}{rrc}
one & two & $\eta_i(\vec k)$ \\
\tableline
 0.1305 & 0.0879 & $1$ \\
-0.5951 &-0.5547 & $\frac{1}{2} (\cos k_x + \cos k_y)$ \\
 0.1636 & 0.1327 & $\cos k_x \cos k_y $ \\
-0.0519 & 0.0132 & $\frac{1}{2} (\cos 2 k_x + \cos 2 k_y)$ \\
-0.1117 &-0.1849 & $\frac{1}{2} (\cos 2k_x \cos k_y + \cos k_x \cos 2k_y)$ \\
 0.0510 & 0.0265 & $\cos 2k_x \cos 2k_y $ \\
\end{tabular}
\end{table}

\begin{figure}
\epsfxsize=3.0in
\epsfbox{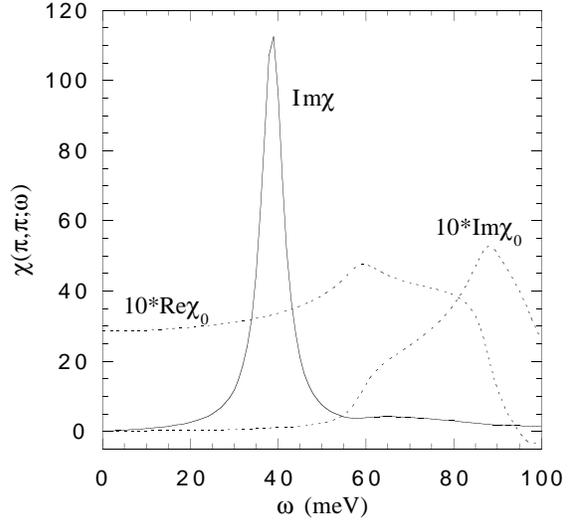}
\vspace{0.5cm}
\label{fig1}
\caption{Im$\chi_0$, Re$\chi_0$, and Im$\chi$ at $q=(\pi,\pi)$ using
dispersion one listed in Table I, with $J$=0.3 eV, $\Gamma$=2.4meV,
$\Delta$=35meV, and $T$=13K.  Im$\chi$ units (Eq.~1) are states/eV/CuO plane,
which applies for all figures.}
\end{figure}

\begin{figure}
\epsfxsize=3.4in
\epsfbox{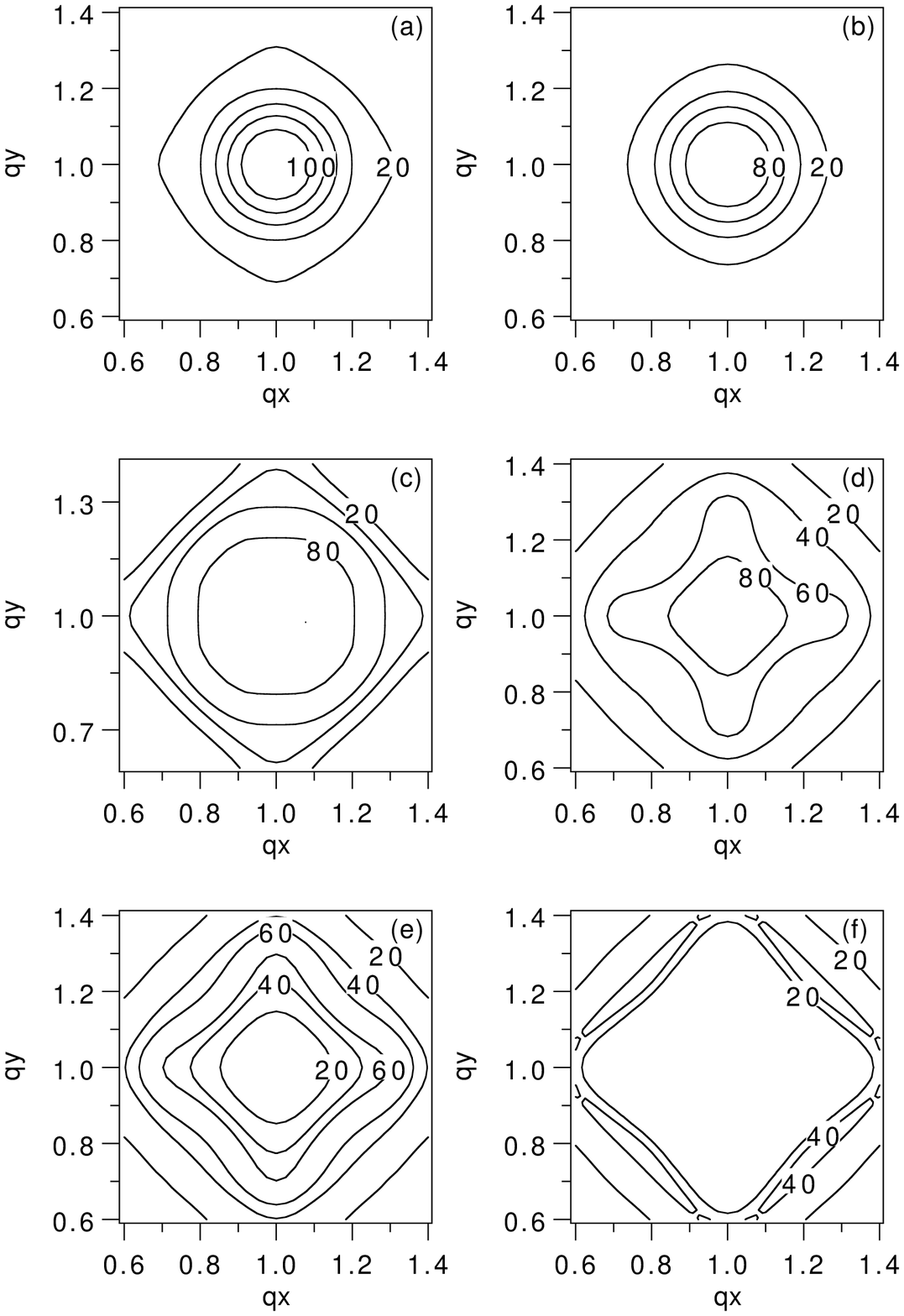}
\vspace{0.5cm}
\label{fig2}
\caption{Dependence of Im$\chi(q_x,q_y)$ at $\omega$=39 meV on
$\epsilon(\pi,0)-\mu$: (a) -34 meV, (b) -10 meV, (c) 0 meV, (d) +5 meV,
(e) +10 meV, and (f) +20 meV (with (a) the value from dispersion one of
Table I).  Same parameters as
Fig.~1, except that for each plot, $J(q)=J$ has been adjusted so
that the maximum in Im$\chi$ is at this $\omega$.  Note that
$qx$ and $qy$ are in units of $\pi$.}
\end{figure}

\begin{figure}
\epsfxsize=3.4in
\epsfbox{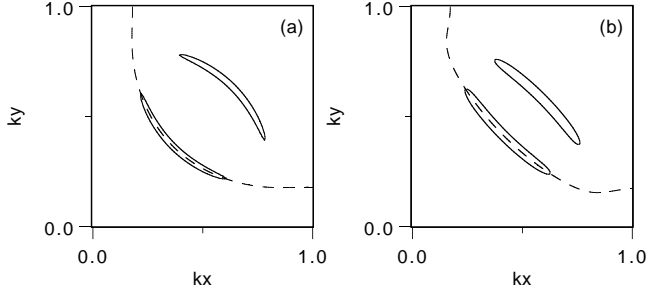}
\vspace{0.5cm}
\label{fig3}
\caption{Fermi surface (dashed curve) and superconducting state energy contours
($E_k,E_{k+Q}=\omega/2$, $\omega$=29 meV), for (a) dispersion one
and (b) dispersion two in Table I.  Note $Q=(1,1)$ in these units.}
\end{figure}

\begin{figure}
\epsfxsize=3.4in
\epsfbox{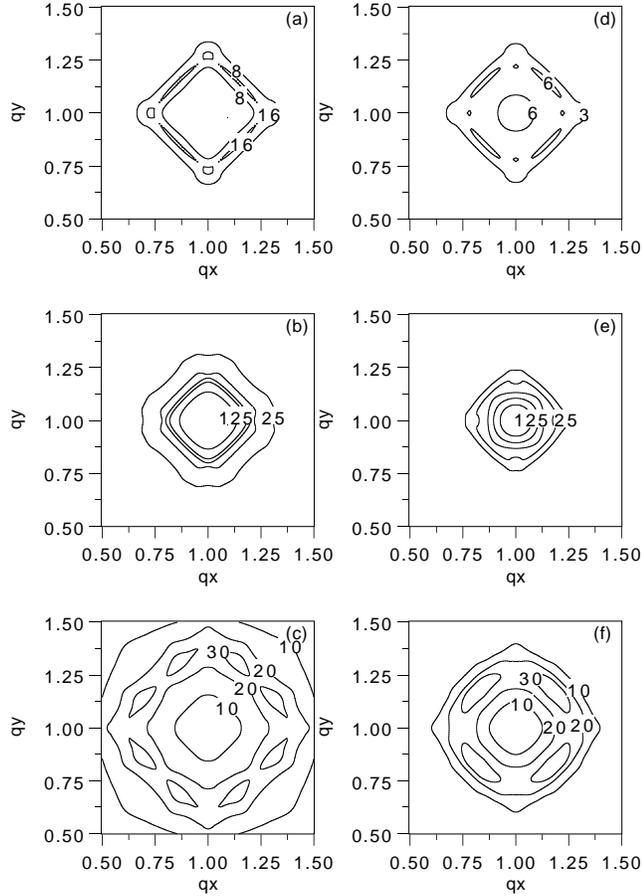}
\vspace{0.5cm}
\label{fig4}
\caption{Im$\chi(q_x,q_y)$ at three energies: (a) 29 meV,
(b) 39 meV (resonance), and (c) 49 meV, for dispersion two of Table I, with
$J$=159 meV ($J(q)=J$).  (d), (e), and (f) are the same, but with $J(q)=J_q$.}
\end{figure}

\begin{figure}
\epsfxsize=3.4in
\epsfbox{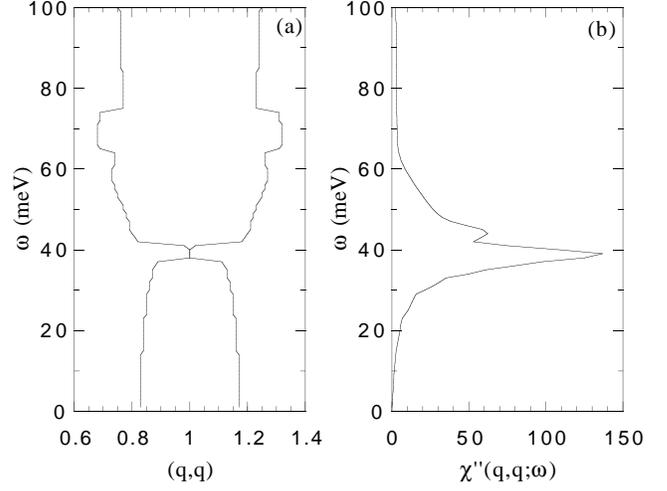}
\vspace{0.5cm}
\label{fig5}
\caption{(a) $q$ vector along $(\pi,\pi)$ direction where Im$\chi(q,q)$
is maximal versus $\omega$. (b) The magnitude of Im$\chi$ in (a) versus
$\omega$.  Same parameters as Fig.~4, with $J(q)=J$.}
\end{figure}

\begin{figure}
\epsfxsize=3.4in
\epsfbox{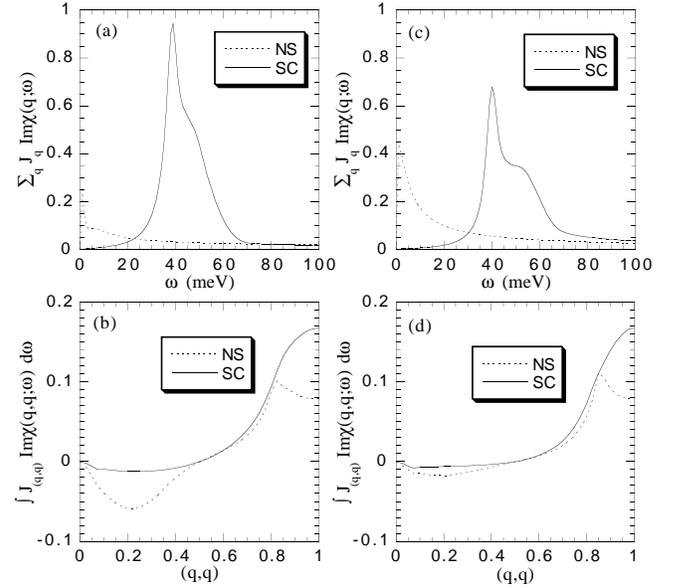}
\vspace{0.5cm}
\label{fig6}
\caption{(a) $\sum_q$ $J_q$Im$\chi(q,\omega)$ and
(b) $\int_0^{100 meV}$ $J_q$Im$\chi(q,\omega)$ for $J(q)=J$.  Same
parameters as in Fig.~5. (c) and (d) are the same, but with $J(q)=J_q$.
NS is the normal state, SC the superconducting state.  The exchange
energy contribution to the condensation energy would be obtained by
integrating (a) or (c) over $\omega$, multiplying by $-3/(2\pi)$,
and subtracting SC from NS.  Note (b) and (d) are in eV units.}
\end{figure}

\begin{figure}
\epsfxsize=3.0in
\epsfbox{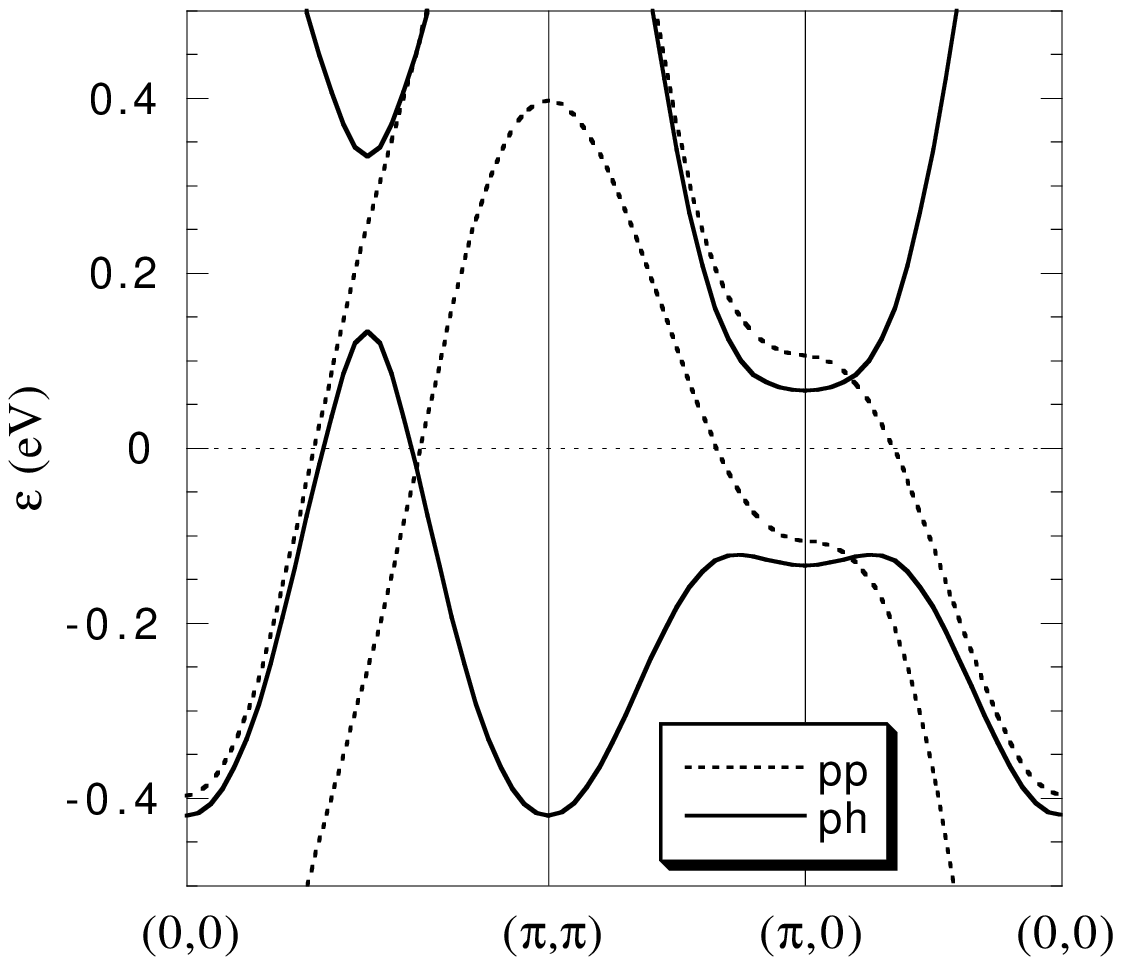}
\vspace{0.5cm}
\label{fig7}
\caption{Solution of particle-hole (ph) secular equation, with diagonal elements
$\epsilon_k, \epsilon_{k+Q}$, and particle-particle (pp) secular equation, with
diagonal elements $\epsilon_k, -\epsilon_{-k+Q}$, with $Q=(\pi,\pi)$.  In
both cases, the off-diagonal elements were taken to be $\Delta_U$=100 meV.
Dispersion one of Table I was employed ($\mu$=0).}
\end{figure}


\begin{references}

\bibitem{FS}
Q. Si, Y. Zha, K. Levin, and J. P. Lu, Phys. Rev. B {\bf 47}, 9055 (1993);
P. Benard, L. Chen, and A.-M. S. Tremblay, Phys. Rev. B {\bf 47}, 15217 (1993);
P. B. Littlewood, J. Zaanen, G. Aeppli, and H. Monien, Phys. Rev. B {\bf 48},
487 (1993).

\bibitem{TRANQ}
J. M. Tranquada, {\it et al.}, Nature {\bf 375}, 561 (1995).

\bibitem{MOOK2}
H. A. Mook, {\it et al.}, Nature {\bf 395}, 580 (1998).

\bibitem{ARAI}
M. Arai, {\it et al.}, Phys. Rev. Lett. {\bf 83}, 608 (1999).

\bibitem{SW}
D. J. Scalapino and S. R. White, Phys. Rev. B {\bf 58}, 8222 (1998).

\bibitem{DZ}
E. Demler and S.-C. Zhang, Nature {\bf 396}, 733 (1998).

\bibitem{MOOK3}
P. Dai, {\it et al.}, Science {\bf 284}, 1344 (1999).

\bibitem{LEVIN}
D. Z. Liu, Y. Zha, and K. Levin, Phys. Rev. Lett. {\bf 75}, 4130 (1995).

\bibitem{SCALAPINO}
N. Bulut and D. J. Scalapino, Phys. Rev. B {\bf 53}, 5149 (1996).

\bibitem{LAVAGNA}
M. Lavagna and G. Stemmann, Phys. Rev. B {\bf 49}, 4235 (1994).

\bibitem{BLEE}
J. Brinckmann and P. A. Lee, Phys. Rev. Lett. {\bf 82}, 2915 (1999).

\bibitem{SCHRIEFFER}
J. R. Schrieffer, {\it Theory of Superconductivity} (Benjamin/Cummings,
Reading, 1964).

\bibitem{ADAM}
A. Kaminski, {\it et al.}, preprint cond-mat/9904390.

\bibitem{VALLA}
T. Valla, {\it et al.}, Science {\bf 285}, 2110 (1999).

\bibitem{RJ}
R. J. Radtke and M. R. Norman, Phys. Rev. B {\bf 50}, 9554 (1994).

\bibitem{NORM95}
M. R. Norman, M. Randeria, H. Ding, and J. C. Campuzano, Phys. Rev. B
{\bf 52}, 615 (1995).

\bibitem{SCHABEL}
M. C. Schabel, {\it et al.}, Phys. Rev. B {\bf 57}, 6090 (1998).

\bibitem{DING96}
H. Ding, {\it et al.}, Phys. Rev. Lett. {\bf 76}, 1533 (1996).

\bibitem{DOUG}
P. Monthoux and D. J. Scalapino, Phys. Rev. Lett. {\bf 72}, 1874 (1994).

\bibitem{FONG}
H. F. Fong, {\it et al.}, Phys. Rev. Lett. {\bf 75}, 316 (1995).

\bibitem{ROSSAT}
J. Rossat-Mignod, {\it et al.}, Physica C {\bf 185-189}, 86 (1991).

\bibitem{MOOK1}
H. A. Mook, {\it et al.}, Phys. Rev. Lett. {\bf 70}, 3490 (1993).

\bibitem{FONG2}
H. F. Fong, {\it et al.}, preprint cond-mat/9910041.

\bibitem{MOOK4}
H. A. Mook, F. Dogan, and B. C. Chakoumakos, preprint cond-mat/9811100.

\bibitem{KEIMER}
H. F. Fong, {\it et al.}, Nature {\bf 398}, 588 (1999).

\bibitem{DESSAU}
Y.-D. Chuang, {\it et al.}, Phys. Rev. Lett. {\bf 83}, 3717 (1999).

\bibitem{AC}
A. Abanov and A. V. Chubukov, Phys. Rev. Lett. {\bf 83}, 1652 (1999).

\bibitem{PRL97}
M. R. Norman, {\it et al.}, Phys. Rev. Lett. {\bf 79}, 3506 (1997);
M. R. Norman and H. Ding, Phys. Rev. B {\bf 57}, R11089 (1998).

\bibitem{HAYDEN}
S. M. Hayden, {\it et al.}, Physica B {\bf 241-243}, 765 (1998).

\bibitem{FENG}
D. L. Feng, {\it et al.}, preprint cond-mat/9908056.

\bibitem{PWA}
P. W. Anderson, {\it The Theory of Superconductivity in the High $T_c$
Cuprates} (Princeton Univ. Pr., Princeton, 1997).

\bibitem{JANKO}
B. Janko, preprint cond-mat/9912073.

\bibitem{IOFFE}
L. B. Ioffe and A. J. Millis, Science {\bf 285}, 1241 (1999).

\bibitem{LEVIN2}
Y. Zha, K. Levin, and Q. M. Si, Phys. Rev. B {\bf 47}, 9124 (1993);
Y.-J. Kao, Q. Si, and K. Levin, preprint cond-mat/9908302.

\bibitem{MESOT}
J. Mesot, {\it et al.}, Phys. Rev. Lett. {\bf 83}, 840 (1999).

\bibitem{LOUIS}
M. Chiao, {\it et al.}, preprint cond-mat/9910367.

\bibitem{SUM}
Formally, it is the structure factor, $S$, which is involved rather than
Im$\chi$, both for the sum rule
and the condensation energy, but the difference is small.

\bibitem{MORR}
Similar overshoots have been reported in previous work; see
D. K. Morr and D. Pines, Phys. Rev. Lett. {\bf 81}, 1086 (1998).

\bibitem{LORAM}
J. W. Loram, K. A. Mirza, and P. F. Freeman, Physica C {\bf 171}, 243 (1990).

\bibitem{COND}
For a very different perspective on this, see M. R. Norman, M. Randeria,
B. Janko, and J. C. Campuzano, preprint cond-mat/9912043.

\bibitem{CHUB}
A. Abanov and A. V. Chubukov, preprint cond-mat/9909385.

\bibitem{DEMLER}
E. Demler and S.-C. Zhang, Phys. Rev. Lett. {\bf 75}, 4126 (1995).

\bibitem{PHENOM}
M. R. Norman, M. Randeria, H. Ding, and J. C. Campuzano, Phys. Rev. B
{\bf 57}, R11093 (1998).

\bibitem{PRL99}
J. C. Campuzano, {\it et al.}, Phys. Rev. Lett. {\bf 83}, 3709 (1999).

\bibitem{KAMPF}
A. P. Kampf and J. R. Schrieffer, Phys. Rev. B {\bf 42}, 7967 (1990).

\bibitem{MARSHALL}
D. S. Marshall, {\it et al.}, Phys. Rev. Lett. {\bf 76}, 4841 (1996).

\bibitem{LAUGHLIN}
R. B. Laughlin, Phys. Rev. Lett. {\bf 79}, 1726 (1997).

\end{references}
\end{document}